\documentclass[aps,prl,amsmath,amssymb,groupedaddress,superscriptaddress,10pt,floatfix,preprint]{revtex4-1}

\usepackage{setspace}   
\usepackage{amsmath}    
\usepackage{graphicx}   
\usepackage{verbatim}   
\usepackage{color}      
\usepackage{subfigure}  
\usepackage{epstopdf}
\setcitestyle{super}
\usepackage{gensymb}
\begin{document}


\title{Ultrafast optically induced ferromagnetic/anti-ferromagnetic phase transition in GdTiO$_3$ from first principles}


\author{Guru Khalsa}
\email[]{guru.khalsa@cornell.edu}
\author{Nicole A. Benedek}
\email[]{nbenedek@cornell.edu}
\affiliation{Department of Materials Science and Engineering, Cornell University, Ithaca, New York 14853, USA}

\date{\today}



\maketitle

\section{Abstract}
Epitaxial strain and chemical substitution have been the workhorses of functional materials design. These \emph{€˜static}€™ techniques have shown immense success in controlling properties in complex oxides through the tuning of subtle structural distortions. Recently, an approach based on the excitation of an infrared active phonon with intense mid-infrared light has created an opportunity for \emph{dynamical} control of structure through special nonlinear coupling to Raman phonons. We use first-principles techniques to show that this approach can dynamically induce a magnetic phase transition from the ferromagnetic ground state to a hidden antiferromagnetic phase in the rare earth titanate GdTiO$_3$ for realistic experimental parameters. We show that a combination of a Jahn-Teller distortion,  Gd displacement, and infrared phonon motion dominate this phase transition with little effect from the octahedral rotations, contrary to conventional wisdom.

\section{Introduction}
The properties of ABO$_3$ perovskites arise not only from the crystal structure generally, they also depend sensitively on small structural distortions, in contrast with most other inorganic materials families. For example, the magnetoresistive properties of perovskite manganites,\cite{hwang95} and metal-insulator transitions in the perovskite nickelate\cite{medarde97,zhou04,catalan08} and titanate families\cite{mochizuki04,pavarini04} are strongly linked with so-called `rotations' of the BO$_6$ octahedra, which are associated with zone-boundary phonons of the cubic perovskite structure. Researchers have learned how to successfully tune this dependence using a variety of techniques, such as epitaxial strain\cite{choi04,lee10} and stabilization,\cite{kim16} pressure,\cite{hiroi93} and doping.\cite{hwang95} Dynamical control of the properties of perovskites has been far more challenging, and has so far been largely based on light-induced electronic excitation or temperature changes.\cite{miyano97,kiryukhin97,zhang16}

Recent experiments have demonstrated the potential for modifying the properties of materials using ultrafast optical pulses to selectively and coherently excite particular phonon modes. One such mechanism involves optical excitation of an IR-active phonon $Q_{IR}$, which produces a displacement of a Raman-active mode $Q_R$ due to a particular anharmonic coupling between the modes of the form $Q_{IR}^2Q_R$ (although the relevant coupling term is not always clear in experimental studies, most theoretical works so far have focused on couplings of the form $Q_{IR}^2Q_R$). This nonlinear phononic effect, and the subtle structural changes it induces, has been invoked to interpret observations of a five-orders-of-magnitude decrease in resistivity in Pr$_{0.7}$Ca$_{0.3}$MnO$_3$,\cite{rini07,subedi14} a material that is insulating at equilibrium at all measured temperatures.\cite{tomioka96,tokura06} Additionally, the charge- and orbitally-ordered phase of La$_{1/2}$Sr$_{3/2}$MnO$_4$ can be destabilized well below the orbital ordering temperature by selective vibrational excitation of an Mn-O stretching mode.\cite{tobey08} In a particularly notable example, the observation of transient superconductivity above the transition temperature in vibrationally excited YBa$_2$Cu$_3$O$_{6.5}$ was also attributed to a nonlinear phononic effect.\cite{mankowsky14} Experiments that exploit this mechanism have, together with first-principles calculations,\cite{subedi14,subedi15,fechner16,juraschek17} produced profound new insights into the structural dynamics of materials out of equilibrium.


Here, we use the family of orthorhombic perovskite titanates to develop a generalized theory of nonlinear phononics, which we use in combination with density functional theory calculations to reveal the existence of a hidden magnetic phase in GdTiO$_3$ accessible under experimentally realistic conditions.  It is generally understood that rotations of the TiO$_6$ octahedra strongly affect the coupling between neighboring Ti-$t_{2g}$ orbitals and therefore control the bandwidth and magnetic superexchange interactions. In the equilibrium phase diagram, the rotation amplitudes increase as the size of the rare earth cation decreases. This change in rotation amplitudes is accompanied by a change in the magnetic ground state, with G-AFM SmTiO$_3$ and FM GdTiO$_3$ straddling the boundary between the AFM and FM parts of the phase diagram. Interestingly, an A-type AFM phase has been speculated to exist between SmTiO$_3$ and GdTiO$_3$, however no stoichiometric system exists between these materials. Although it may seem that the A-AFM phase could be obtained by judiciously tuning octahedral rotation angles, Mochizuki and Imada\cite{mochizuki04} point out that the Jahn-Teller mode plays a critical role in stabilizing the A-AFM phase (Pavarini and co-workers had a similar insight\cite{pavarini05}). In particular, the A-AFM phase requires the combination of a ``strong'' Jahn-Teller distortion (as in GdTiO$_3$) and ``moderate'' octahedral rotation distortion (as in SmTiO$_3$). However, the Jahn-Teller distortion is coupled to the rotations and weakens rapidly as the rotations decrease. There is no known way to independently control the relative magnitudes of these distortions. Indeed, as far as we are aware, neither bulk chemical techniques (solid solutions) nor strain engineering have been able to experimentally stabilize the A-AFM phase in any perovskite titanate.

We demonstrate that the nonlinear phononics mechanism can be used to essentially ``switch off'' the net magnetization in GdTiO$_3$ by inducing an A-AFM phase. Although we find that the A-AFM phase can be realized dynamically in bulk GdTiO$_3$, the required peak electric fields are quite high. However, we show that a modest amount of epitaxial strain can position GdTiO$_3$ closer to the phase boundary, such that the A-AFM phase is induced at low peak electric fields of the optical pulse. We also elucidate the origin of the magnetic transition in terms of the structural distortions that accompany the optical excitation. Surprisingly, although the octahedral rotation distortions are affected by optical excitation, their amplitudes change very little from their equilibrium values and their effect on the magnetic configuration energy is negligible. Instead, we find that it is primarily the Jahn-Teller mode that plays the key role in switching off the net magnetization in GdTiO$_3$, and dynamically inducing the A-AFM phase. \emph{These are precisely the conditions under which Mochizuki and Imada speculated that the A-AFM phase could be realized.} Our results demonstrate that the nonlinear phononics mechanism can be leveraged to modulate the complex interplay between structural distortions in perovskites, and their electronic and magnetic properties. We also shed light on the structural dynamics of perovskites out of equilibrium, a regime in which these materials have been little explored.

\section{Results}
\subsection{Theory}
\textit{Two-Mode Model.--} An expansion of the lattice energy clarifies the nonlinear phononics process for a centrosymmetric crystal in which the IR and Raman modes have different symmetry. For simplicity we first consider a model involving only a single IR mode and a single Raman mode (this model is similar to that presented in Refs. \citenum{forst11} and \citenum{subedi14}). The lattice energy $U$ is defined by 

    \begin{equation}
    	U = \frac{1}{2}\omega_{IR}^2 Q_{IR}^2 +
        	\frac{1}{2}\omega_{R}^2 Q_{R}^2 +
            A Q_{IR}^2 Q_{R} +
            \Delta\vec{P} \cdot \vec{E}(t),
        \label{eqn:Energy_Simple}
    \end{equation}

\noindent
where $\omega_{IR(R)}$ and $Q_{IR(R)}$ are, respectively, the frequency and amplitude of the IR (Raman) phonon and $A$ is a coupling coefficient. The last term contains the polarization change in the crystal, defined as $\Delta\vec{P} = \tilde{Z}^* Q_{IR}$, where $\tilde{Z}^*$, known as the mode effective charge,\cite{zhao02} describes the strength of the coupling between the excited IR phonon and the electric field of the light pulse ($\vec{E}$) (we ignore the effects of the oscillating IR mode on the polarization). Note that since $Q_{IR}^2$ is invariant under all symmetry operations of the crystal, in order for $A$ to be non-zero, $Q_R$ must also be invariant under all symmetry operations of the crystal. 
By convention, we label this totally symmetric mode $A_g$. (Nonlinear coupling of the form $Q_{IR}Q_{R}^2$, linear in the IR mode displacement and quadratic in Raman mode displacement, are not allowed because of inversion symmetry (taking $Q_{IR} \rightarrow -Q_{IR}$). In noncentrosymmetric crystals, some IR phonons are fully symmetric. In this case, these terms are allowed.)


We desire a large, \emph{static} displacement of the Raman mode that is long-lived on the timescale of the IR mode. Taking derivatives of Eqn. \ref{eqn:Energy_Simple} with respect to the IR and Raman amplitudes for a fixed electric field direction,

	\begin{align} \label{eqn:EOM_Simple_IR}
    \ddot{Q}_{IR} &= -\omega_{IR}^2 Q_{IR} - 2 A Q_{IR} Q_{R}-\tilde{Z}^*E(t),\\
	\ddot{Q}_{R}  &= -\omega_{R}^2 Q_{R} - A Q_{IR}^2,
    \label{eqn:EOM_Simple_R}
    \end{align}
\noindent
reveals that this depends on three materials-specific criteria: 1) Strong coupling between the excited IR mode and the incoming light pulse (large $\tilde{Z}^\ast$\cite{gonze97}), 2) Strong coupling between the excited IR mode and the Raman mode (large $A$) and; 3) The frequency of the IR phonon should be higher than that of the Raman mode, such that the Raman mode does not oscillate within the period of the IR mode. Eqn. \ref{eqn:EOM_Simple_R} also shows that as long as the IR phonon is ringing, there will be a unidirectional force on the Raman mode.

Eqns. \ref{eqn:Energy_Simple} -- \ref{eqn:EOM_Simple_R} provide useful insight into the basic physical process exploited in the nonlinear phononics mechanism, however real materials contain many IR and Raman phonons. In general, when an IR mode is optically excited, \emph{all} symmetry-allowed modes will contribute to the dynamical response. The two-mode model ignores the dynamic coupling between Raman modes and will provide a poor description of the structural dynamics when the coupling between these modes is strong. Additional complications arise when more than one IR mode is excited by the incoming light pulse, as shown in Figure \ref{manymodes}. As the oscillator strength of the second mode increases, the structural dynamics becomes increasingly complicated, and the two-mode model will again fail to provide a correct description. Ultimately, without knowing the relevant nonlinear coupling coefficients explicitly, it is not possible to predict \textit{a priori} which modes will be strongly coupled or have a large contribution to the dynamical response. Hence, all IR and Raman phonons should be treated on equal footing. We present a model below that does just that.

\textit{Many-Mode Model.--} In order to make the problem concrete and tractable, here we focus on the case where light is polarized along the principal crystallographic axes and parallel with the particular IR phonon we wish to excite. Additionally, we use the perovskite $Pbnm$ crystal structure as a starting point, without loss of generality, because it is the most stable crystal structure for more than half of all perovskites, including GdTiO$_3$ (extension of this model to arbitrary crystal systems -- with or without inversion symmetry -- is straightforward, but cumbersome).
 
The orthorhombic \emph{Pbnm} structure has 20 atoms per primitive cell and therefore 60 zone-centered phonons. There are 25 IR-active phonons: 9 polarized along each of the \textit{a}- and \textit{b}-axes ([100] and [010]), and 7 polarized along the \textit{c}-axis ([001]). Additionally, there are 7 Raman-active phonons of $A_g$ symmetry that are accessible through the nonlinear phononics process. Expanded to third order, Eqn. \ref{eqn:Energy_Simple} becomes, 
\begin{equation}
\begin{aligned}
U & = \frac{1}{2} \sum_i \omega_i^2 Q_i^2 
+ \frac{1}{2} \sum_\alpha \omega_\alpha^2 Q_\alpha^2 
+ \sum_{i\alpha} A_{ii\alpha} Q_i^2 Q_\alpha \\
&+ \sum_{ij\alpha} B_{ij\alpha} Q_i Q_j Q_\alpha 
+ \sum_{\alpha \beta \gamma} C_{\alpha \beta \gamma} Q_\alpha Q_\beta Q_\gamma 
+ \Delta\vec{P} \cdot \vec{E}(t).
\end{aligned}
\label{eqn:Energy_Expanded}
\end{equation}

\noindent
We use Latin indices $(i,j,...)$ for IR-active phonons and Greek indices $(\alpha,\beta,...)$ for Raman-active phonons. Again, $A_{ii\alpha}$ leads to a unidirectional displacement of mode $\alpha$ but now multiple Raman modes may be displaced by selective excitation of IR mode $i$. The $B_{ij\alpha}$ terms allow for coupling between two different IR modes and a Raman mode. Juraschek and co-workers explored this coupling term theoretically for the case of two IR phonons polarized along different directions, where a coupling to non-$A_g$ Raman phonons is induced.\cite{juraschek17} Nova et al. have explored this term experimentally by simultaneously exciting two IR phonons with different polarizations in ErFeO$_3$.\cite{nova_effective_2017} This term also couples two IR phonons with parallel polarization, again displacing an $A_g$ Raman mode. The $C_{\alpha \beta \gamma}$ term couples three displaced Raman modes. Since multiple IR modes are now included the change in the polarization must also be expanded as $\Delta\vec{P} = \sum_i \tilde{Z_i}^* Q_i$.

Taking derivatives of Eqn. \ref{eqn:Energy_Expanded} gives the equations of motion for the nonlinear phononics process to third order:
\begin{equation}
\begin{aligned}
    \ddot{Q}_{i} = &-\omega_{i}^2 Q_{IR} 
    - 2 \sum_\alpha A_{ii\alpha} Q_i Q_\alpha \\
    & -2\sum_{j\alpha} B_{ij\alpha} Q_j Q_\alpha  - \tilde{Z_i}^*E(t)
\end{aligned}
\label{eqn:EOM_Expanded_IR}
\end{equation} 

\begin{equation}
\begin{aligned}
	\ddot{Q}_{\alpha}  &= -\omega_{\alpha}^2 Q_{\alpha} 
    - \sum_{i}A_{ii\alpha} Q_{i}^2 \\
    & -\sum_{ij} B_{ij\alpha} Q_i Q_j
    - 3 \sum_{\beta \gamma} C_{\alpha \beta \gamma} Q_\beta Q_\gamma
\end{aligned}
\label{eqn:EOM_Expanded_R}
\end{equation}
\noindent We describe below how we integrate the many-mode model with density functional theory calculations to obtain a first-principles description of the structural dynamical response of a given material to optical excitation.


\textit{Integrating the Many-Mode Model with First-Principles Calculations.--} Figure \ref{workflow} shows the workflow we have implemented for obtaining the required inputs to the many-mode model, and for exploring the dynamical response of materials induced through the nonlinear phononics mechanism. A number of different steps are involved:
\begin{itemize}
\item{\textbf{Relax the crystal structure of interest} The atomic positions and lattice parameters of the given material are fully relaxed using density functional theory. Forces must be tightly converged to ensure phonon properties are accurate.}
\item{\textbf{Compute phonon properties and Born charges} Phonon frequencies, eigenvectors and Born effective charges are calculated using density functional perturbation theory. The eigenvectors and Born charges are required for calculation of the mode effective charges, which appear in Eqns. \ref{eqn:EOM_Expanded_IR} and \ref{eqn:EOM_Expanded_R}, along with the phonon frequencies.}
\item{\textbf{Model lattice contribution to the dielectric function} The phonon properties and Born charges are also used to model the lattice dielectric function. This is not strictly required for simulating the dynamical response, however it is helpful for connecting our model with experimental information, such as the reflectivity and zero-frequency dielectric response.}
\item{\textbf{Compute third-order coupling terms} The third-order nonlinear coupling coefficients $A_{ii\alpha}$, $B_{ij\alpha}$ and $C_{\alpha \beta \gamma}$ are calculated with density functional theory using finite differences. Details are provided in the Methods section.}
\item{\textbf{Simulate dynamical response} A particular IR mode is selected for optical excitation, and the dynamical response of the crystal is simulated by numerically solving Eqns. \ref{eqn:EOM_Expanded_IR} and \ref{eqn:EOM_Expanded_R}. The optical pulse characteristics (timescale, shape, and peak electric field) are described in the Methods section.}
\item{\textbf{Explore consequences for functional properties} The many-mode model outputs a set of IR and Raman amplitudes, which are added back to the relaxed crystal structure to give a snapshot of the dynamical response of the material to optical excitation. The consequences for the band gap, magnetic configurations, and other properties can then be explored using density functional theory.}
\end{itemize}

\subsection{Magnetic phase transition in GdTiO$_3$}
GdTiO$_3$ adopts the distorted $Pbnm$ crystal structure with a ferromagnetic alignment of the Ti 3d$^1$ spins below $\sim$30 K. As mentioned above, GdTiO$_3$ contains a number of IR-active modes and we found it was possible to position the system \emph{both} closer to and further away from the FM -- A-AFM phase boundary, depending on which mode we chose to excite. We compared the magnetic configuration energy of the A-AFM, G-AFM and C-AFM states with the FM state for excitation of various IR modes and identified a 72.4 meV phonon polarized along the crystallographic  $c$-axis ([001] direction) as being most likely to induce the desired magnetic phase transition. For all the results presented below, we checked that the material remains insulating. Figure \ref{Figure_4} shows the difference in the magnetic configuration energy between the FM and A-AFM states as a function of peak electric field (we also checked the energy difference between the FM phase and the other AFM orders but the A-AFM phase remains closest in energy to FM over the entire peak electric field range). The optically induced change in magnetic energy results from both the oscillating IR phonons (due to a biquadratic spin-phonon coupling) and the unidirectionally displaced Raman phonons. This is most clearly understood by expanding the out-of-plane ($c$-axis) Heisenberg exchange constant, $J_c$, in both the oscillating excited mode $Q_{IR,0}$ and the total unidirectionally displaced Raman mode amplitudes $Q_{R,0}$. By symmetry alone, $J_c \rightarrow J_c(1+aQ_{R,0}+bQ_{IR,0}^2)$, where $a$ and $b$ are negative coefficients in this case. The interpretation is that the oscillating IR phonon induces a change in the effective exchange constants because it appears as $Q_{IR,0}^2$. That is, although the time average of $Q_{IR}$ is zero -- the average amplitude does not change -- the time average of $Q_{IR}^2$ is \emph{not} zero. This is essentially the same physics that produces the unidirectional displacement of the Raman phonons, since the IR mode also appears as $Q_{IR}^2$ in the relevant coupling term in Equations \ref{eqn:Energy_Simple} and \ref{eqn:Energy_Expanded}.  This effect of the IR phonon on magnetism has not been addressed previously in the nonlinear phononics literature, however our results suggest that coupling between the spins and IR phonons cannot be ignored. 

Figure \ref{Figure_4} shows that although the system approaches the A-AFM state with increasing peak electric field, the FM state remains stable over the entire electric field range. The magnetic transition could be induced by simply increasing the peak electric field, however it is desirable to keep the peak electric field as low as possible, in order to minimize sample damage and heating. We therefore need some way to position GdTiO$_3$ closer to the FM -- A-AFM phase boundary.


Recent work has shown how strain engineering can tune La$_{(2/3)}$Ca$_{(1/3)}$MnO$_3$ into a charge-ordered insulating phase with extreme photo-susceptibility.\cite{zhang16}  Ultrafast optical excitation of the strained material induces a transition to a long-lived, hidden metallic phase. Likewise, we calculated the energy difference between the A-AFM, G-AFM, C-AFM and FM states as a function of both tensile and compressive biaxial strain (the strain was applied in the plane perpendicular to the $c$-axis). Figure \ref{Figure_5}a shows that both tensile and compressive strain decrease the energy of all of the AFM phases relative to the FM state, however the A-AFM phase again remains closest in energy to the FM state. In fact, epitaxial strain \emph{by itself} can induce the A-AFM phase in GdTiO$_3$. However, the strains required (tensile strains greater than 2\% and compressive strains greater than 2.5\%) are quite large and likely unrealizable in real GdTiO$_3$ thin-films. Nonetheless, calculation of the relevant exchange coefficients and magnetic $T_C$ does show how strain moves GdTiO$_3$ closer to the FM -- A-AFM phase boundary. The difference between the FM and A-AFM phases corresponds to a change from FM to AFM ordering along the $c$-axis of the crystal. The out-of-plane exchange coefficient ($J_c$) tracks this energy difference well, as Figure \ref{Figure_5}b shows. As the in-plane lattice constants increase, the in-plane ($ab$-plane) exchange coefficients ($J_{ab}$) decrease. This leads to a steady decrease in the critical $T_C$ for tensile strain (Figure \ref{Figure_5}c). Because of the sharp decrease in the out-of-plane exchange coefficient ($J_c$) with tensile strain we look for optically induced magnetic switching at $+1.5\%$ tensile strain, where the critical temperature is still large and $J_{ab}$ is still greater than zero. (+1.5\% tensile strain corresponds approximately to growth on GdScO$_3$.)\cite{choi04,uecker2006} Previous theoretical work has also explored strain as a possible strategy for stabilizing the A-AFM phase in rare-earth titanates (and specifically GdTiO$_3$) although much larger values of strain were typically required.\cite{huang2013,weng2014,yang2014,huang2015}


Figure \ref{Figure_6}(a) now shows that excitation of the same IR-active phonon (now at 70.8 meV in the strained material) pushes GdTiO$_3$ across the phase boundary, stabilizing the A-AFM phase at lower peak electric fields than in the unstrained system. We find that the combined peak IR and Raman response stabilizes the A-AFM phase for $E_0 > 2.5$ MV/cm. The average unidirectional Raman response alone requires $E_0 > 4.5$ MV/cm to induce this change, a still modest peak electric field value. These results suggest that \emph{a previously unidentified magnetic phase of the rare earth titanates may be reached dynamically by the nonlinear phononics process.}


What is the structural origin of the switch from the FM to the A-AFM phase? To answer this question, we use the symmetry-adapted modes of the parent cubic $Pm\bar{3}m$ phase as a basis to decompose the nonequilibrium structural changes into contributions from different distortions.(See Supplementary Information) Because the Raman displacement is quasi-static, we focus on this aspect of the structural changes. We group the components of the Raman modes into three categories, which have been highlighted in the titanate literature because of their effect on electronic and magnetic properties in the equilibrium phase diagram: two octahedral rotation distortions (transforming like the irreducible representations $R_4^+$, for the $a^-a^-c^0$ rotation in Glazer notation, and $M_3^+$ for the in-phase $a^0a^0c^+$ rotation), three distortions that change Ti-O bond lengths ($R_5^+$:O , $X_5^+$:O, and $M_2^+$:O represent distortions of the octahedral oxygen environment) and modes involving displacements of the Gd cations ($R_5^+$:Gd and $X_5^+$:Gd). Each $A_g$ phonon is a linear combination of all three kinds of structural distortion, although some modes may be dominated by a particular type of distortion.


The two most commonly discussed structural distortions in the perovskite literature are the out-of-phase $R_4^+$:O and in-phase $M_3^+$:O octahedral rotations. As the schematic phase diagrams in Figure \ref{Figure_7} show, the amplitudes of these distortions increase in moving from La to Y in the rare-earth titanate series ATiO$_3$ (A={La,Nd,Sm,Gd,Y}); there is a concomitant change in magnetic order from G-AFM in LaTiO$_3$ to FM in YTiO$_3$. This change in magnetic order is usually understood as a response to changes in the Ti-O-Ti bond angles (which are directly affected by the rotation amplitudes) and subsequent changes in the magnetic superexchange interactions. However, if we try to interpret our results in terms of this picture, we find something unexpected. First, let us consider our strain results and focus on small strains $\pm$1\% around the bulk value, the region that is most likely to be applicable to thin-film experiments. Figure \ref{Figure_5}a shows that small tensile strains favor the A-AFM phase, whereas small compressive strains further stabilize the FM ground state. Now, how does this compare with the correlation between rotation angles and magnetism in bulk? The schematic phase diagrams in Figure \ref{Figure_7} show that increasing the $R_4^+$ and $M_3^+$ rotation angles favors the FM phase, whereas decreasing them favors the AFM phase. However, Figure \ref{Figure_7} also shows that compressive strain changes the $R_4^+$ and $M_3^+$ rotation angles such that the system is pushed towards the AFM-FM phase boundary, whereas small tensile strains push the system further into the FM region (red shaded region). This is the opposite behavior to what we would expect based on the results in Figure \ref{Figure_5}. Quite astonishingly, if we now consider the dynamical response (blue shaded region), we see that as the peak electric field increases, the $R_4^+$ angle changes in such a way that the system moves away from the AFM-FM phase boundary. Moreover, the change in rotation angle across the whole range of strains and peak electric fields is very small, barely more than 1$^\circ$. Similar results hold for the $M_3^+$ rotation (Figure \ref{Figure_7}b), where the change in rotation angle is even smaller. Clearly, considering only changes in rotation angles cannot explain the transition to A-AFM order under optical excitation in GdTiO$_3$. (The strain results are essentially a volume effect but we defer further discussions to a forthcoming publication so we can focus here on the dynamical response.)

If changes in rotation angles do not induce a change in magnetic order in GdTiO$_3$ through the nonlinear phononics mechanism, then which structural distortion is responsible? Figure \ref{Figure_8}a shows the percentage change in the symmetry-adapted mode distortion amplitudes as a result of optical excitation (the same plot showing absolute changes can be found in the Supplementary Information). Positive changes indicate that the distortion increases in magnitude relative to the equilibrium structure, whereas negative changes indicate that the distortion decreases in magnitude.  The  $M_2^+$ Jahn-Teller mode undergoes the largest change by far, increasing in magnitude with respect to the equilibrium structure (this corresponds to a 158\% change in the Jahn-Teller amplitude at for a peak electric field of 7 MV/cm). The $R_5^+$ mode, which changes the O and Gd environments, also increases in magnitude. In contrast, the two octahedral rotation modes, $R_4^+$ and $M_3^+$, change very little. \emph{Contrary to conventional wisdom, our results suggest that judicious tuning of octahedral rotation angles is unlikely to produce the desired A-AFM phase under experimentally realizable conditions. Instead, the Jahn-Teller mode is playing the key role in the magnetic switching process.} To test this insight, we re-calculated the energy difference between the FM and A-AFM phases while selectively returning specific sets of distortions to their equilibrium values. Figure \ref{Figure_8}(b) shows that when all modes are returned to their equilibrium values \emph{except} for the two rotation modes, the FM phase is stable over the entire range of peak electric fields we consider.  When all modes are returned to their equilibrium values except the rotations and those involving displacements of the Gd ions, there is a crossover from the FM to the A-AFM phase at high peak electric fields. Inclusion of the $M_2^+$ Jahn-Teller mode pushes the crossover to lower peak electric fields.

\section{Discussion}
Our results demonstrate that the nonlinear phononics mechanism can be exploited to dynamically stabilize the A-AFM phase of GdTiO$_3$ in the manner suggested by Mochizuki and Imada -- by modulating the amplitude of the Jahn-Teller mode while changing the rotations very little, a combination of conditions that is extremely difficult to realize in either bulk or thin-film GdTiO$_3$. The ``indirectness'' of the nonlinear phononics process -- light is used to excite a particular phonon mode, which then produces a displacement of another set of modes through anharmonic coupling -- is sometimes seen as a disadvantage of the mechanism, since ions are only displaced from their equilibrium values by very small amounts. In this case however, the indirect nature of the mechanism is probably precisely what allows the stabilization of the A-AFM phase. Octahedral rotations are very low-energy distortions, in that they lower the energy of the cubic perovskite phase by a significant amount (over 1 eV in some cases) and have very large amplitudes in the distorted $Pbnm$ structure. Large amounts of energy, supplied through epitaxial strain, for example, are therefore required to change their amplitudes. In contrast, the amplitude of the Jahn-Teller mode is typically orders of magnitude smaller. We speculate that the nonlinear phononics mechanism supplies just enough energy to induce changes in the Jahn-Teller mode in GdTiO$_3$, but not enough to significantly affect the octahedral rotations. Of course, the dynamical response of other perovskites may be different; the structures of perovskites out of equilibrium are only beginning to be explored in detail, and so much more work is needed in this area. 

There is some experimental support for our findings. Zhang and co-workers\cite{zhang13} grew very thin-films of GdTiO$_3$, such that the octahedral rotations are suppressed. However, they found that ferromagnetism persisted, even in the absence of rotation distortions, and posit that the magnetism is controlled by ``the narrow bandwidth, exchange and orbital ordering''. Regardless of the mechanism, clearly octahedral rotations are not as essential to the magnetic properties of GdTiO$_3$ as has perhaps been assumed. Experimental confirmation of the dynamical magnetic response can in principle be achieved by time-resolved Faraday or Kerr probes\cite{kampfrath} referenced to the infrared lattice excitation, while corresponding birefringence measurements can detect changes in the Jahn-Teller mode amplitude.

We conclude by emphasizing that we are emphatically \emph{not} claiming that octahedral rotations are never important in giving rise to the electronic and magnetic properties of perovskites. There is a large and well-established theoretical and experimental literature unambiguously showing the various ways in which these distortions affect the properties of perovskites. However, in the case of the rare earth titanates at least, it appears that octahedral rotations merely ``set the stage'' for the magnetic properties, while much smaller structural distortions (such as the Jahn-Teller mode) work together to induce the actual magnetic ground state. 


\section{Methods} 
\noindent
\textbf{First-principles calculations.}
First-principles calculations were performed using projector augmented wave potentials and the PBEsol + U exchange-correlation functional, as implemented in VASP. On-site Coulomb and exchange parameters of $U_{Ti,d}$ = 5 eV, and $J_{Ti,d}$ = 0.64 eV were used\cite{okamoto_lattice_2006} (we found no qualitative change in our results for reasonable variations of $U$). We use a 600 eV plane wave cut-off and a $6\times 6\times 4$ Monkhorst-Pack grid. These choices produced good agreement with available experimental structural data for ATiO$_3$ (A={La,Nd,Sm,Gd,Y}). \cite{maclean_crystal_1979}(See Supplemenatary Figure 3 for detailed comparison of GdTiO$_3$.) For GdTiO$_3$ we find the lattice constants $a=5.372$ \AA, $b=5.768$ \AA, $c=7.646$ \AA, in good agreement with available high-temperature experimental values $a=5.393$ \AA, $b=5.691$ \AA, $c=7.664$ \AA. Phonons and Born charges were calculated using density functional perturbation theory within VASP and used to calculate dynamical mode effective charges ($\tilde{Z_i}^*$) and to model the frequency-dependent dielectric response. \cite{gonze_dynamical_1997}

\noindent
\textbf{Magnetic exchange and critical temperature.}
Magnetic exchange constants were calculated by fitting the A-, C-, G-AFM, and FM configuration energies to a Heisenberg model $E_{magnetic}=-\sum_{\left<i,j\right>_\gamma} J_\gamma \bf{S_i \cdot S_j}$, where $\left<i,j\right>_\gamma$ labels nearest neighbors for in-plane ($\gamma=ab$) and out-of-plane ($\gamma=c$) spins. We note that including both Raman and IR phonons in the Heisenberg model results in an effective exchange coupling $J_\gamma \rightarrow J_{\gamma}(1+ a_{\gamma} Q_{R,0}+ b_{\gamma} Q_{IR,0}^2)$, with both $a_{c}Q_{R,0}$ and $b_{c}Q_{IR,0}^2$ reaching negative values, inducing the FM to A-AFM phase transition. Here $Q_{IR,0}$ is the excited IR phonon and $Q_{R,0}$ is the total induced Raman distortion - see Fig. \ref{Figure_8}a.) The critical temperature $T_C$ is evaluated from the spin-$1/2$ molecular field theory result with total exchange field of $4J_ab+2J_c $ per spin. 

\noindent
\textbf{Strain response of GdTiO$_3$.} 
Strain was considered for both common growth modes of \textit{Pbnm} perovskites. The first corresponds to growth along the $c$-axis and maintains the \textit{Pbnm} space group (\# 62). The second corresponds to growth perpendicular to the $c$-axis. This lowers the symmetry to $P2_1/m$ (\#11). We find that for tensile strain on orthorhombic substrates, the \textit{Pbnm} phase is generally preferred. In the specific case addressed in the main text, +1.5\% strain corresponds approximatly to growth on GdScO$_3$.

\noindent
\textbf{Calculation of non-equilibrium force constants.}
To calculate the non-equilibrium force constants, we use a finite-difference approach where force constant matrices ($\Phi$) are calculated for each $A_g$ phonon, displaced about equilibrium. The non-equilibrium force constant matrices are then used to calculate the nonlinear coupling coefficients between the $A_g$ and the IR phonons ($A_{ii\alpha}$, $B_{ij\alpha}$, $C_{\alpha \beta \gamma}$) by projecting the force constant matrices into the basis of the equilibrium structure. For example, the real-space coupling between $A_g$ Raman mode $\alpha$, and IR modes $i$ and $j$ is found by $\left(\Phi_{ij}(u_\alpha)-\Phi_{ij}(-u_\alpha)\right)/2u_\alpha$. Here, $u_\alpha$ is the real-space eigendisplacement amplitude associated with phonon $Q_\alpha$. In $Pbnm$ this approach requires 15 phonon calculations: 1 for the equilibrium structure, and two increments for each of the 7 $A_g$ modes. All third-order terms that couple to $Q_\alpha$ are found in this way. We therefore have \emph{all} necessary third-order coupling terms for the nonlinear phononics process. This approach is significantly less computationally burdensome than finite-difference techniques that calculate all third-order terms, or frozen-phonon techniques that require converged meshes in phonon amplitude. In $Pbnm$ the frozen phonon technique would require 175 two-dimensional meshes to calculate $A_{ii\alpha}$, 651 three-dimensional meshes to calculate $B_{ij\alpha}$, and 84 three-dimensional meshes to calculate all $C_{\alpha \beta \gamma}$. Regardless, we have tested our computational approach against the frozen phonon method for multiple randomly selected $A_{ii\alpha}$, $B_{ij\alpha}$, and $C_{\alpha \beta \gamma}$ and found good agreement. We note that this computational technique may also be used for coupling to non-$A_g$ Raman modes. In that case $u_\alpha$ labels the real-space eignedisplacement amplitude of the non-$A_g$ mode of interest.

\noindent
\textbf{Optical pulse characteristics.}
Phonon frequencies, dynamical mode effective charges, non-linear coupling coefficients, and optical pulse characteristics are fed into Eqns. \ref{eqn:EOM_Expanded_IR} and \ref{eqn:EOM_Expanded_R}, which are numerically simulated to find the dynamical response. We include all IR phonons polarized parallel to the incident light as well as all $A_g$ phonons in our simulations.  We fix the timescale $\tau$ (full-width at half-maximum) of a Gaussian pulse to be 300 fs and vary the peak-electric field ($E_0$) from 1 MV/cm to 7 MV/cm for carrier frequencies on resonance with the desired IR phonon (for a Gaussian pulse the extrinsic quantity $(E_0 \tau)^2$ affects the Raman subspace; the electric field range has been chosen consistent with experimentally reported values.\cite{rini07,mankowsky_ultrafast_2017})

\section{Data Availability Statement}
Data is available upon reasonable request from the corresponding author.

\section{Acknowledgments}
This work was supported by the National Science Foundation. NAB was supported by DMR-1550347. Initial work on this project by GK was supported by DMR-1550347, and subsequently by the Cornell Center for Materials Research with funding from the National Science Foundation MRSEC program (DMR-1719875). This work used the Extreme Science and Engineering Discovery Environment (XSEDE) (through allocation DMR-160052), which is supported by National Science Foundation grant number ACI-1548562. We thank Farhan Rana, Jeffrey Moses, Craig Fennie and Gregory Fuchs for helpful discussions.

\section{Competing Interests}
The authors declare no competing financial interests.

\section{Author Contributions}
The authors contributed equally to this work.

\section{Supplementary Information}
Supplementary information is available at the npj Quantum Materials' website.

\newpage
\section{References} 

\newpage
\section{Figure Legends} 

\begin{figure}[h!]
\centering\includegraphics[width=5.375in]{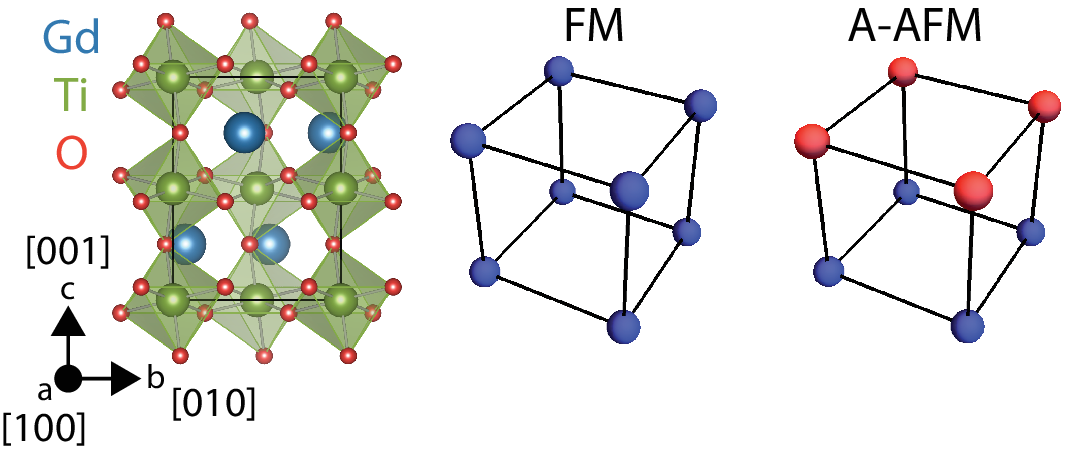}
\caption{\label{structures}\textbf{Crystal and magnetic structures of GdTiO$_3$} Orthorhombic structure of GdTiO$_3$ in the $Pbnm$ setting and schematic showing ferromagnetic (FM) and A-type antiferromagnetic (A-AFM) magnetic configurations for the Ti 3d$^1$ spins. The red and blue spheres correspond to different spin directions.}
\end{figure}

\begin{figure}[h!]
\includegraphics[width=5.375in]{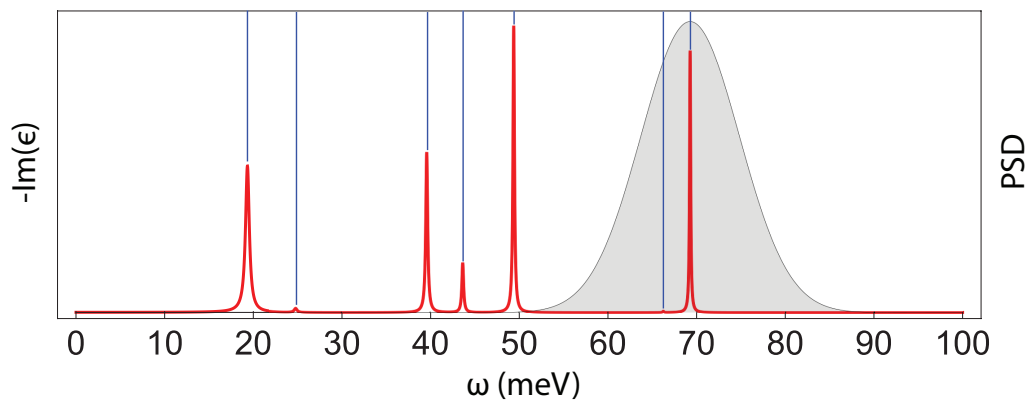}
\caption{\label{manymodes}\textbf{Simultaneous excitation of multiple IR modes} Schematic of the imaginary part of the dielectric function (left axis) with superimposed 300 fs Gaussian pulse (right axis, power spectral density). The horizontal blue lines highlight the positions of the IR modes, which are all polarized along the same crystallographic axis. Even though the pulse is tuned to be resonant with a particular IR mode, because the pulse has a finite width, it will also excite other IR modes that are close in frequency to the target phonon.}
\end{figure}

\begin{figure}[h!]
\includegraphics[width=5.375in]{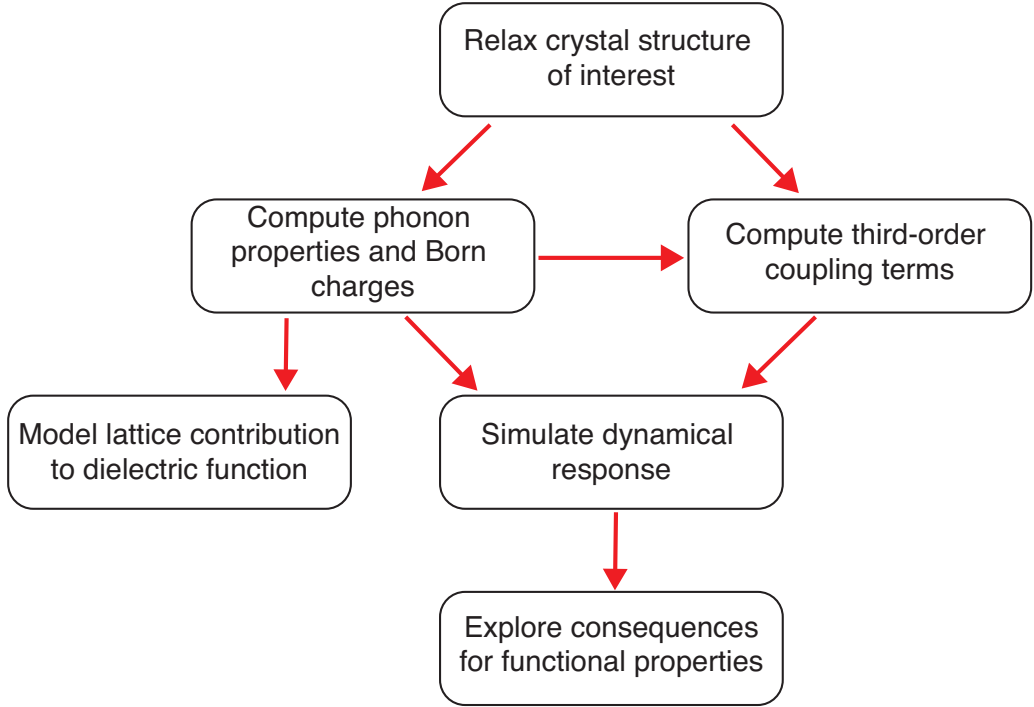}
\caption{\label{workflow}\textbf{Practical implementation of many-mode model} Workflow illustrating the calculations required and steps involved in simulating the dynamical response of a material to optical excitation using the many-mode model.}
\end{figure}

\begin{figure}[h!]
\includegraphics[height=7.5cm]{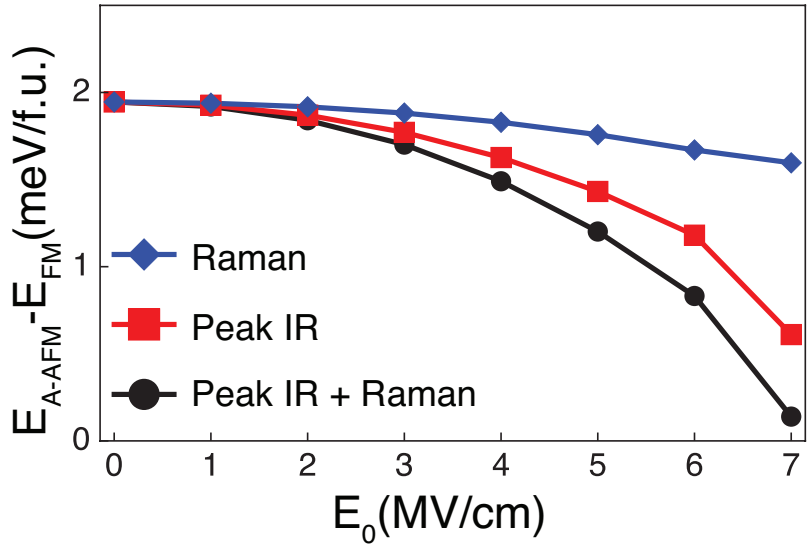}
\caption{\label{Figure_4}\textbf{Phonon-induced change in magnetism in bulk GdTiO$_3$} Simulated energy difference between the A-AFM and FM phases as a function of peak electric field for 72.4 meV, [001] polarized, IR phonon. A positive energy difference indicates that the FM phase is more stable than A-AFM.}
\end{figure}

\begin{figure}[h!]
\includegraphics[height=15.5cm]{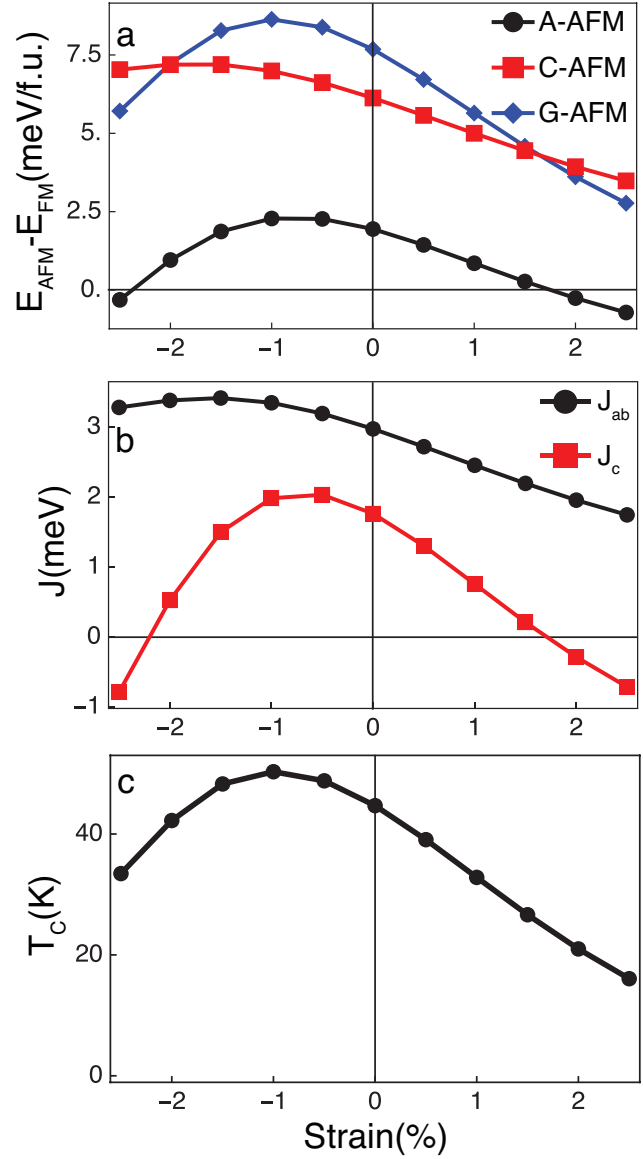}
\caption{\label{Figure_5}\textbf{Effect of epitaxial strain on magnetism in GdTiO$_3$} (a) Magnetic configuration energy differences relative to the FM state. Again, positive energy differences indicate that the FM state is more stable than a particular AFM phase. Negative strains correspond to the compressive region, whereas positive strains are tensile. (b) In-plane ($J_{ab}$) and out-of-plane ($J_c$) Heisenberg exchange parameters ($J > 0$ corresponds to FM order) and (c) Weiss mean field critical temperature as a function of in-plane strain.}
\end{figure}

\begin{figure}[h!]
\includegraphics[height=7.5cm]{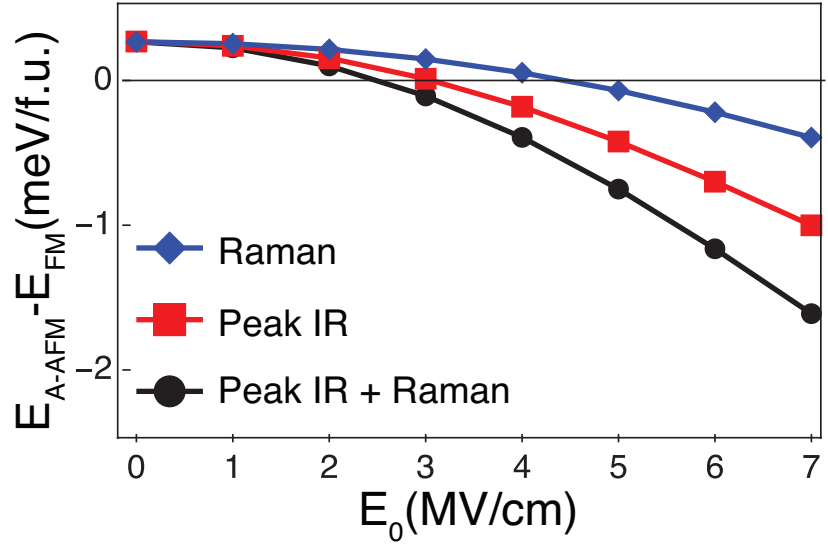}
\caption{\label{Figure_6}\textbf{Phonon-induced change in magnetism in +1.5\% strained GdTiO$_3$.} Simulated energy difference between the A-AFM and FM phases as a function of peak electric field for 70.8 meV, [001] polarized, IR phonon. This mechanism now stabilizes the A-AFM phase at experimentally realizable peak electric fields.}
\end{figure}

\begin{figure}[h!]
\includegraphics[width=3.375in]{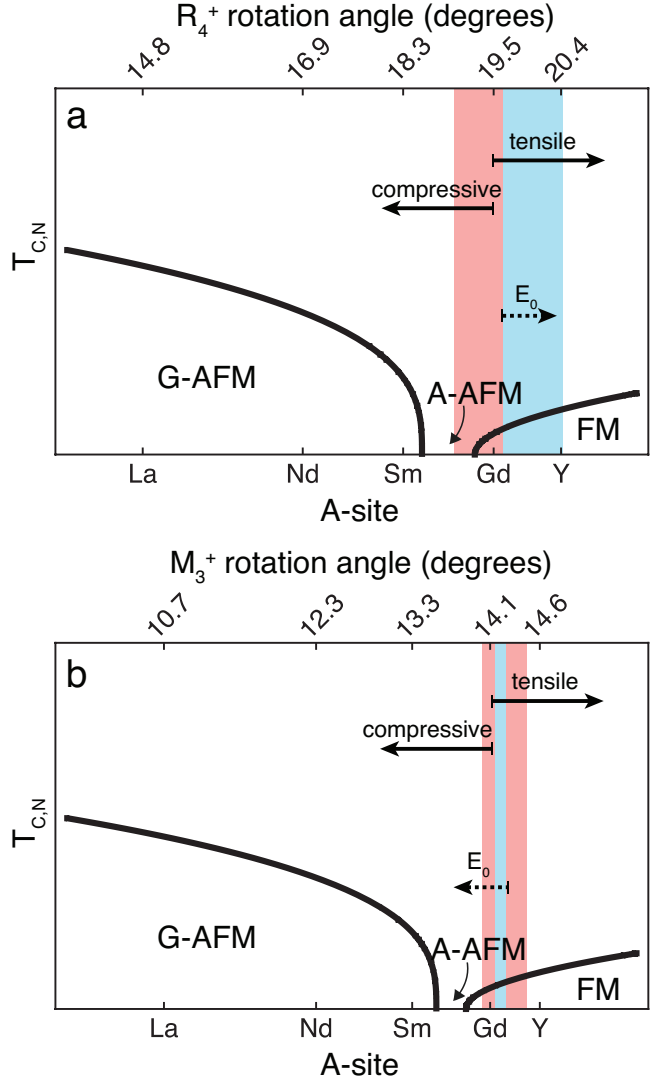}
\caption{\label{Figure_7}\textbf{Titanate magnetic phase diagram versus rotation angle} The (a) out-of-plane rotation ($R_4^+$) and (b) in-plane rotation angles ($M_3^+$) accessed by epitaxial strain (red) and the nonlinear phononics approach in +1.5\% strained GdTiO$_3$ (blue) cannot account for the change in magnetism to the A-AFM phase. The lower horizontal axis shows the A-site while the upper horizontal axis shows a least squares fit to the rotation angle from the simulated structures in the corresponding magnetic phase. Phase boundaries have been drawn schematically with magnetic phases identified. The predicted A-AFM phase is shown in the purple region. Arrows showing the qualitative change in rotation angle have been included for the compressive/tensile strain (solid) and peak electric field $E_0$ (dashed).}
\end{figure}

\begin{figure}[h!]
\includegraphics[width=3.375in]{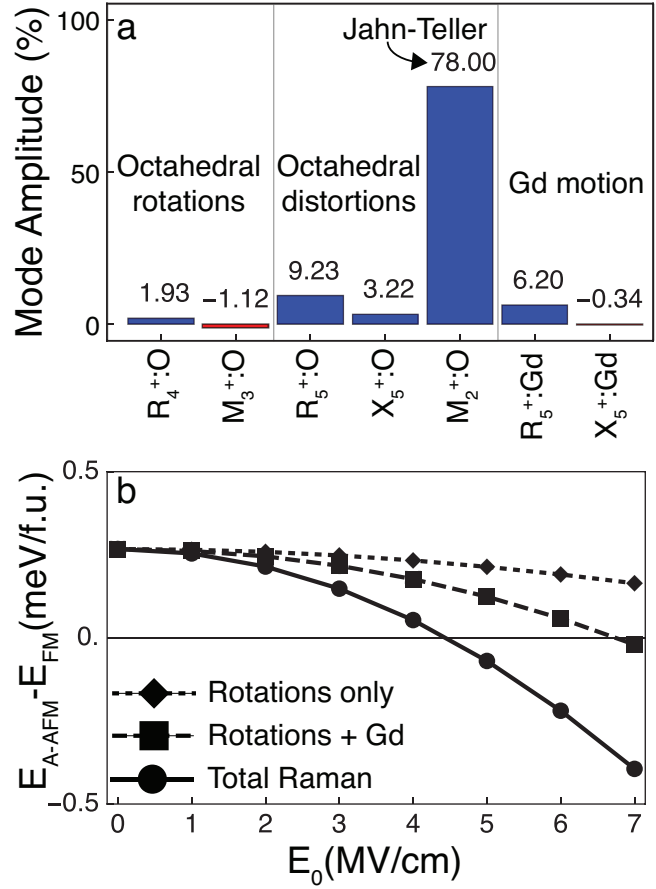}
\caption{\label{Figure_8}\textbf{Structural contributions to the magnetic phase change } (a) Percent change in the symmetry-adapted mode amplitude for the Raman component of the nonequilibrium strained GdTiO$_3$ structure. Blue (red) corresponds to an increase (decrease) in amplitude. Note that the percentages sum to 100\%. (b) Nonlinear phononics change in the magnetic energy difference between the A-AFM and FM phases with octahedral rotations only (dashed), rotations and octahedral distortions (dotted), and total Raman response (solid).}
\end{figure}

\end{document}